\documentclass[]{article}
\usepackage{graphicx} 
\usepackage{amsmath}
\usepackage{mathtools}
\usepackage{amssymb}
\usepackage[svgnames]{xcolor}
\usepackage{braket}
\usepackage{graphicx}
\usepackage[utf8]{inputenc}
\usepackage{tikz}
\usetikzlibrary{backgrounds,calc,positioning}
\graphicspath{ {./images/} }
\usepackage{natbib}
\usepackage[nottoc]{tocbibind}
\usepackage{authblk}

\title{Dirac-von Neumann Type Axiomatic Structure for Classical Electromagnetism }
\author[]{Daniel W. Piasecki \\ dpiasec@ncsu.edu \\ https://orcid.org/0000-0002-3399-115X}
\affil[]{Department of Physics, North Carolina State University}

\begin{document}
\maketitle


\section*{Abstract}
We demonstrate the existence of a complex Hilbert Space with Hermitian operators for calculations in \textit{classical} electromagnetism that parallels the Hilbert Space of quantum mechanics. The axioms of this classical theory are the so-called Dirac-von Neumann axioms, however, with classical potentials in place of the wavefunction and the indeterministic collapse postulate removed. This approach lets us derive a variety of fundamental expressions for electromagnetism using minimal mathematics and a calculation sequence well-known for traditional quantum mechanics. We also demonstrate the existence of the wave commutation relationship $[\hat{x},\hat{k}]=i$, which is a unique classical analogue to the canonical commutator $[\hat{x},\hat{p}]=i\hbar$. The difference between classical and quantum mechanics lies in the presence of $\hbar$. The noncommutativity of observables for a classical theory simply reflects its wavenature. A classical analogue of the Heisenberg Uncertainty Principle is developed for electromagnetic waves, and its implications discussed. Further comparisons between electromagnetism, Koopman-von Neumann-Sudarshan (KvNS) classical mechanics (for point particles), and quantum mechanics are made. Finally, supplementing the analysis presented, we additionally demonstrate an elegant, completely relativistic version of Feynman's proof of Maxwell's equations \citep{Dyson}. Unlike what \citet{Dyson}
indicated, there is no need for Galilean relativity for the proof to work. This fits parsimoniously with our usage of classical Lie commutators for electromagnetism.


\section{Classical and Quantum Hilbert Spaces}

Quantum theory is a unique theory based on a complex Hilbert Space with Hermitian operators acting on the vector kets to produce eigenvalues. The superficially strange mathematics of quantum theory has for a long time haunted many great physicists. There have been many attempts made to make quantum mathematics appear in form to resemble the classical functions of position, momentum, and time used in Newtonian physics, in order to avoid the complex Hilbert space with vectors living in an infinite number of dimension \citep{kvnwaves,MauroPhDThesis}. Some hoped this would make quantum theory more interpretable or less mysterious seeming, however, the Hilbert Space formalism could not be avoided. 

Koopman-von Neumann-Sudarshan (KvNS) mechanics was an early approach to go the other way. Instead of representing quantum mechanics in terms of more familiar representations, it showed that one can take classical mechanics and represent it inside a complex Hilbert space as well. Koopman, von Neumann, and Sudarshan were able to take the same postulates of quantum mechanics and establish a purely classical theory \citep{koopman,vN,Sudarshan, kvnwaves, MauroPhDThesis,ODM,Koopmanwavefunctions,McCauletal,PiaseckiThesis}. KvNS mechanics, like standard quantum mechanics, uses Hermitian operators acting on state kets to calculate expectation values of observables \citep{kvnwaves,MauroPhDThesis,ODM,McCauletal}. It has a classical wavefunction ket $\ket{\psi}$ and a Born Rule to calculate probabilities. Also like with quantum theory, it has the curious property of collapse of the waveform once a classical measurement is made, although particular details differ \citep{MauroPhDThesis}. One could say that this makes classical mechanics a hidden variable theory of quantum mechanics \citep{Sudarshan}. \citet{ODM} was able to show that the only difference between classical theory and the quantum theory is in the choice of the position-momentum commutator. For relativistic and nonrelativistic classical mechanics, position and momentum commute: 

$$[\hat{x},\hat{p}]=0$$
Using this fact and the Koopman algebra \citep{Sudarshan,ODM,relODM,McCauletal,PiaseckiThesis}, one is able to compute the Liouville equation for classical probability densities. Operational Dynamic Modeling (ODM) takes the KvNS and quantum formalism and merges them together to explore theoretical questions of interest \citep{ODM,wignerphasespace,relODM}. 

Here we present another classical Hilbert Space approach for electromagnetic fields. Like KvNS mechanics, it will also contain Hermitian operators acting on kets and the eigenvalue problem. However, unlike both quantum theory and KvNS, we will maintain electromagnetism's deterministic flavor, and there will not be any probabilistic/collapse interpretation for it (Born Rule for the wavefunction analogues). This approach will be shown to be useful, giving us a simple, quantum-like way to make relevant expansions for electromagnetism. It might also be pedagogically useful, as it provides a simple set of rules to develop in a rigorous manner typical expressions commonly used for electromagnetic configurations (see also \citealt{Hanson_and_Yakovlev}). Surprising consequences of this approach are explored, including the fact that the commutator for electromagnetism is not the classical KvNS commutator, but:

$$[\hat{x},\hat{k}]=i$$
The implication for KvNS mechanics and quantum mechanics will be discussed.

The paper aims to reformulate the well-established theory of classical electromagnetism in terms of the \textit{mathematical} language of Hilbert Spaces, operators, and commutators in a fashion paralleling quantum mechanics.  While quantum mechanics famously uses these tools, it does not monopolize them \citep{Stoler,Hanson_and_Yakovlev,classical_light_and_QM}. As demonstrated here, a fully deterministic theory of classical fields can be derived with the same mathematical language and a similar set of axioms to the well-known Dirac-von Neumann axioms. 

Although previous works have drawn attention to the operator and Hilbert space structure of classical electromagnetism (e.g., \citealt{Stoler,other_HS_EM_theory,Hanson_and_Yakovlev,CE16,classical_light_and_QM}), this paper differs from these previous works in several significant ways. For one, previous work has not demonstrated axiomatically how classical electromagnetism can be based on a set of axioms similar to the Dirac-von Neumann axioms (the same set of axioms except the Born rule for $\ket{A_0}$ and $\ket{\boldsymbol{A}}$ fields and collapse of the waveform, which is where quantum indeterminism arises). This makes it very simple to make comparisons between different Hilbert space theories of both quantum and classical mechanics, as everything is based in a similar set of postulates. We highlight some similarities and differences between quantum and classical concepts in this work. 
Two, we demonstrate that the noncommutativity of certain classical field observables arises directly from the wave nature of the fields (eq. \ref{eq:correspondence}), independent of any quantum assumptions. This structure emerges not from quantization, but as a mathematical consequence of wave propagation itself. Whether the wave is classical or quantum is secondary to the fact that wave-like systems naturally support noncommuting operators. This observation allows us to formulate a classical analogue of the canonical commutator and, by extension, a deterministic analogue of the Heisenberg Uncertainty Principle. While noncommuting operators have been discussed in classical contexts—for instance, in the algebraic formulation of Koopman classical mechanics \citep[]{Morgan}—such treatments typically develop formal frameworks in which noncommuting structures may be introduced, rather than deriving explicit commutator relations for directly interpretable observables within classical electromagnetism.
Three, in light of the operator structure of the classical field and inspired by the approach taken in ODM, we demonstrate that Feynman's controversial proof of the homogeneous Maxwell equations can be made wholly relativistic. ODM utilizes the relativistic canonical commutator to derive the Dirac equation from basic assumptions \citep{OpenDirac,relODM}, and this same commutator can be shown to lead to Maxwell's homogeneous equations \citep{Dyson}. Feynman's original proof was plagued with nonrelativistic equations leading to relativistic ones, a problem we completely eliminate here (Appendix B). Past work has focused on using the Lagrange formalism or a Poisson structure to fix Feynman's proof, but in the same style of the original work \citep{Dyson}, we present for the first time a relativistic classical/quantum Lie commutator proof.  
The Hilbert Space axioms of electromagnetism give us a self-consistent view. In addition to the commutator structure and foundational axioms discussed so far, this paper also briefly examines Green’s function in both quantum and electromagnetic Hilbert Spaces (further discussed in \citealt{Hanson_and_Yakovlev}), along with other useful mathematical information for this approach provided throughout (see Appendix A). The content of this theory is still purely that of \textit{classical} electromagnetism, and is not meant as a substitute for QED.

Feynman taught us that it is important to have a variety of mathematical tools and approaches to any theoretical problem, as where one set of tools may fail, another theory might capture underlying physical processes (for instance, we are reminded of Feynman's allegory of the Mayan astronomer, which highlights how having one model that calculates well some physical property of interest may not be enough to elucidate deeper understanding in physics). In a similar vein, Maxwell stated ``it is a good thing to have two ways of looking at a subject, and to admit that there \textit{are} two ways of looking at it" \citep{maxwell_book}. Heading Maxwell's and Feynman's advice, we explore the Hilbert Space for electromagnetic potentials and fields. This is a proof-of-concept paper for this approach, which we plan to further develop in future works.

\section{A Complex Hilbert Space for Electromagnetism}

\subsection{Axioms of Theory}
KvNS classical mechanics and quantum mechanics are built on the same axioms, which allows for the unified investigative framework of Operational Dynamical Modeling \citep{ODM,relODM,Koopmanwavefunctions,McCauletal}. We summarize the so-called Dirac-von Neumann axioms for both classical KvNS and quantum mechanics as follows \citep{dirac,von_N_textbook,Shankar, Sakurai,PiaseckiThesis}:

\begin{enumerate}
    \item The 
    wavefunction ket $\ket{\psi}$ (a vector in complex Hilbert Space) describes the state of the system
    \item For observable $\zeta$, there is an associated Hermitian operator $\hat{\zeta}$ which obeys the eigenvalue problem $\hat{\zeta}\ket{\zeta} = \zeta\ket{\zeta}$, where $\zeta$ is the value seen by measurement. Two common observables are position $\hat{x}$ and momentum $\hat{p}$, which obey (in one-dimension)

    $$\hat{x}\ket{x}=x\ket{x}$$
    $$\hat{p}\ket{p}=p\ket{p}$$
    \item Born Rule: The probability density for making any measurement for observable $\zeta$ is given by 
    $$\rho_\zeta = \braket{\psi|\zeta}\braket{\zeta|\psi}$$
    Upon measurement, the state of the system collapses from $\ket{\psi}$ to $\ket{\zeta}$.
    
    \item The state space of a composite system is the tensor product of the subsystem’s state spaces, $\mathcal{H} = \mathcal{H}_1 \otimes \mathcal{H}_2 \otimes ...$. 

    \item Unitary time evolution: \begin{equation}
    \ket{\psi(t)}=\hat{U}(t-t')\ket{\psi(t')}
\end{equation}
where
\begin{equation}\label{eq:propogator}
    \hat{U}=\exp(-i\hat{\Omega}(t-t'))
\end{equation}
with
\begin{equation*}
    \hat{\Omega}\ket{\omega}=\omega\ket{\omega}
\end{equation*}
\begin{equation*}
    \hat{\mathbb{I}} = \int d\omega ~ \ket{\omega}\bra{\omega}
\end{equation*}
The operator $\hat{\Omega}$ is Hermitian and here the infinitesimal generator of time evolution of classical and quantum wavefunctions. In the quantum case, $\hat{\Omega}$ is the Hamiltonian $\hat{H}$, and in the KvNS case, it is the Louivillian $\hat{L}$, sometimes referred to as the Koopman generator \citep{Koopmanwavefunctions, PiaseckiThesis}. Planck's constant does not appear in the classical case.
\end{enumerate}
The above lacks terminology of quantum or classical, \textit{because it is the set of postulates that encompasses both} \citep{koopman,vN,Sudarshan,ODM,relODM}. 

For the classical electromagnetic theory, we will begin with a related set of axioms as our foundation, but for elements of the four-potential. In the last century, physicists have realized the importance of the electromagnetic four-vector $A_\mu =(V, A_x, A_y, A_z)$. It has been argued by some that the four-potential is in ways more fundamental than the electric and magnetic fields, which act on the level of force, whereas the four-potential exists on the level of potentials and momentum \citep[etc.]{ABeffect,AB1961,ABFeynman,Calkin1,Calkin2,Konopinski,Calkin3,Gingras,mead, ABextended,feynmanpotentialsfirst2020}. Here, the potentials will be treated as analogous to the wavefunction.  Any square integrable function can be represented as an element of a $L^2$ Hilbert Space \citep{byron_and_fuller,hilbertoperators,gallone}. For any square integrable components $A_\mu(\zeta, t)$, we can represent them as an abstract vector ket $\ket{A_\mu (t)}$ existing in the Hilbert Space, signifying the component state of the potential $A_\mu$ (where $\mu = 0,1,2,3$). The space 
\begin{equation}\label{eq:space_L}
    {L}^2(M,d\zeta) = \{\psi: M \rightarrow \mathbb{C} | \int d\zeta ~\psi^*\psi = N < \infty\},
\end{equation}
completely analogous to the quantum case \citep{dirac,Shankar, hilbertoperators,gallone}, will be the space of electromagnetic potential \textit{scalar} elements. 

In the standard framework, linear operators $\hat{\zeta}: \mathcal{H} \rightarrow \mathcal{H}$ are defined as linear mappings on the complex Hilbert space \citep[pp. 17]{hilbertoperators}. The eigenvalue $\zeta$ of linear operator $\hat{\zeta}$ is defined as a complex number characterized by the condition that $\hat{\zeta}-\zeta\hat{I}$ is non-injective \citep[pp. 25]{hilbertoperators}. Corresponding eigenvectors $\ket{\zeta} \in \mathcal{H}$ satisfy the eigenvalue expression $\hat{\zeta}\ket{\zeta}=\zeta\ket{\zeta}$ \citep[pp. 25]{hilbertoperators}. 

For the electromagnetic Hilbert Space, we propose the following axioms:
\begin{enumerate}
    \item The kets $\ket{A_0}$ and $\ket{\boldsymbol{A}}$ (vectors in complex Hilbert Space) describe the state of the four-potential elements $V$ and $\boldsymbol{A}$, respectively.
    \newline\newline
    Notation: 
    $$\ket{A_\mu} = ~~\begin{cases} 
           ~\ket{A_0} ~~for ~ scalar ~ potential \\
          ~\ket{\boldsymbol{A}} ~~for~vector~potential 
       \end{cases}$$
       \newline\newline
       This is possible due to the fact that any square integrable function is a member of a $L^2$ Hilbert Space.
    
    \item For classical position and wavenumber, there exists an associated Hermitian operator $\hat{x}$ for position and $\hat{k}$ for wavenumber which obey the one-dimensional eigenvalue problems

    $$\hat{x}\ket{x}=x\ket{x}$$
    $$\hat{k}\ket{k}=k\ket{k}$$
    The observable $x$ represents a position along a physically real wave spread out in space, for example. We consider the wavenumber $k$ a classical observable since the wavelength is in principle classically measurable. The eigenvalues of the classical Hermitian operators are experimentally verifiable quantities. 
    \item The state space of a composite system is the tensor product of the subsystem’s state spaces, $\mathcal{H} = \mathcal{H}_1 \otimes \mathcal{H}_2 \otimes ...$
    \newline\newline
    The introduction of the Hilbert space tensor product structure here serves as an abstract mathematical tool to combine distinct Hilbert spaces in a deterministic setting. Unlike in KvNS and QM, it is not used to represent joint probability amplitudes.
    \item Unitary time evolution:
    \begin{equation}
    \ket{A_\mu(t)}=\hat{U}(t-t')\ket{A_\mu(t')}
\end{equation}
where
\begin{equation}\label{eq:propogator}
    \hat{U}=\exp(-i\hat{\Omega}(t-t'))
\end{equation}
with
\begin{equation*}
    \hat{\Omega}\ket{\omega}=\omega\ket{\omega}
\end{equation*}
\begin{equation*}
    \hat{\mathbb{I}} = \int d\omega ~ \ket{\omega}\bra{\omega}
\end{equation*}
The operator $\hat{\Omega}$ is Hermitian and here the infinitesimal generator of time evolution of classical fields, corresponding to the wave frequency. Planck's constant does not appear in the classical case. 
Eq. \ref{eq:propogator}
is precisely the same mathematical object as the familiar phase factor $e^{-i\omega(k) t}$ used to represent monochromatic wave evolution in classical electromagnetic theory (e.g., \citealt{jackson}, \citealt{EMprop} eq. 2.15 and details therein, etc.). This further validates the interpretation of $\hat{U}$ as a classical wave evolution operator within Hilbert Space.
\end{enumerate}

Postulate 1 is in direct comparison to quantum mechanics, where the abstract ket $\ket{\psi}$ describes the state of the quantum system \citep{dirac, von_N_textbook,Shankar, Sakurai}. For example, since $\ket{A_0}$ is the abstract representation of the electric potential $V$, whose position basis would give us the familiar scalar valued field utilizing the Hilbert Space inner product on a continuous basis:

\begin{equation}\label{eq:closureV}
    V(\boldsymbol{x},t) = \braket{\boldsymbol{x}|A_0(t)},
\end{equation}
Because $\boldsymbol{A}(\boldsymbol{x},t)$ is a Euclidean vector with three components, we can represent it as the tensor product of a polarization state and scalar amplitude portion, or
\begin{equation}\label{eq:vector_potential}
\ket{\boldsymbol{A}}=\ket{\boldsymbol{n}}\ket{\Phi}.
\end{equation}
$\ket{\boldsymbol{n}}$ lives in $\mathbb{R}^3$ and the \textit{magnitude of the vector potential} $\ket{\Phi}$ lives in the infinite-dimensional functional Hilbert space (i.e., an element of the $L^2$ space, eq. \ref{eq:space_L}), so that $\ket{\boldsymbol{A}}$ technically lives in the tensor product of the two spaces, $\mathcal{H}_1 \otimes \mathcal{H}_2$. We can always expand $\ket{\boldsymbol{A}}$ across any Euclidean basis $\{\ket{e_i}\}$, for example:

$$\ket{\boldsymbol{A}}=\ket{e_1}\braket{e_1|\boldsymbol{n}}\ket{\Phi}+\ket{e_2}\braket{e_2|\boldsymbol{n}}\ket{\Phi}+\ket{e_3}\braket{e_3|\boldsymbol{n}}\ket{\Phi}$$
where $\braket{e_i|\boldsymbol{A}}$ is our notation for the 3-Euclidean vector dot product in Hilbert Space. Then, we can simply write that closure with the position basis gives us

\begin{equation}\label{eq:closureA}
    \boldsymbol{A}(\boldsymbol{x},t) = \braket{\boldsymbol{x}|\boldsymbol{A}(t)}.
\end{equation}

It is important to stress that this is a theory on the level of the gauge fields $V$ and $\boldsymbol{A}$. As such, to have a coherent theory on the level of the gauge fields, we still need to impose the usual gauge conditions \citep{jackson}, which we do here just like in usual standard electromagnetic theory. Because gauge conditions (Lorenz, Coulomb, etc.) are imposed, this is a gauge symmetry
preserving theory, just like standard electromagnetism. 

An observant reader will notice the Born Rule has been removed from set of electromagnetic postulates.
Unlike the Hilbert Space for quantum mechanics, we do not impose any probabilistic or collapse interpretation on the mathematics. Any square integrable function can have a representation in a complex Hilbert space \citep{byron_and_fuller,hilbertoperators, gallone}. In the context of quantum mechanics, the property of square integrability is exploited to normalize a wavefunction so that sensible probabilities can be extracted (from eq. \ref{eq:space_L}).

For the electromagnetic case, we are merely interested in the property that the integral in eq. \ref{eq:space_L} is finite for different objects of interest, and will not be mapping the finite amplitude into a probability density. Therefore, a large difference between the classical wave and quantum theory is that the quantum theory has a probabilistic interpretation, but the classical wave theory is deterministic due to lack of such an imposition. The mathematical structure, however, is otherwise identical. 

We will see that the second postulate of the electromagnetic Hilbert Space theory leads to a commutator relationship analogous to that of the canonical commutator. As we will show, it will lead to a classical Heisenberg Uncertainty Principle when the electromagnetic amplitude is normalizable \citep[]{classicaluncertaintyprinciple, Mansuripur}. The third postulate, borrowed from the other Hilbert Space theories, will be necessary in carrying out certain calculations, as we will demonstrate. The fourth postulate is utilized for time-dependent Green's Operator (section 6). A Venn diagram summarizes the relationship between KvNS classical point mass mechanics, quantum theory, and this electromagnetic theory (Figure 1). 
\begin{figure}
    \centering
    \begin{tikzpicture}
   \begin{scope} [fill opacity = .4]
    \draw[draw = black] (-2.5,1) circle (5);
    \draw[draw = black] (2.5,1) circle (5);
    \draw[draw = black] (0,-4) circle (5);
    \node at (-4.5,6.5) {\LARGE\textbf{KvNS Mechanics}};
    \node at (4.5,3) {\textbf{3N coordinate space}};
    \node at (5,1) {\Large\textbf{Indeterministic}};
    \node at (3,-2.5) {\Large\textbf{$[\hat{x},\hat{k}]=i$}};
    \node at (-4.5,1.5) {\LARGE\textbf{$[\hat{x},\hat{k}]=0$}};
    \node at (4,6.5) {\LARGE\textbf{Quantum Mechanics}};
    \node at (-3,-2.5) {\large\textbf{Deterministic}};
    \node at (0,0) {\Large\textbf{Hilbert Space $\mathcal{H}$}};
    \node at (0,3.5) {\Large\textbf{Born Rule}};
    \node at (0,2.5) {\textbf{Collapse of Waveform}};
    \node at (0,-6) {\Large\textbf{3-Euclidean space (restricted)}};
    \node at (0,-9.5) {\LARGE\textbf{Electromagnetism}};
    \end{scope}
\end{tikzpicture}
    \caption{Venn Diagram of relationship between Koopman-von Neumann-Sudarshan mechanics, standard quantum mechanics, and the electromagnteic potential Hilbert Space theory.}
    \label{fig:enter-label}
\end{figure}
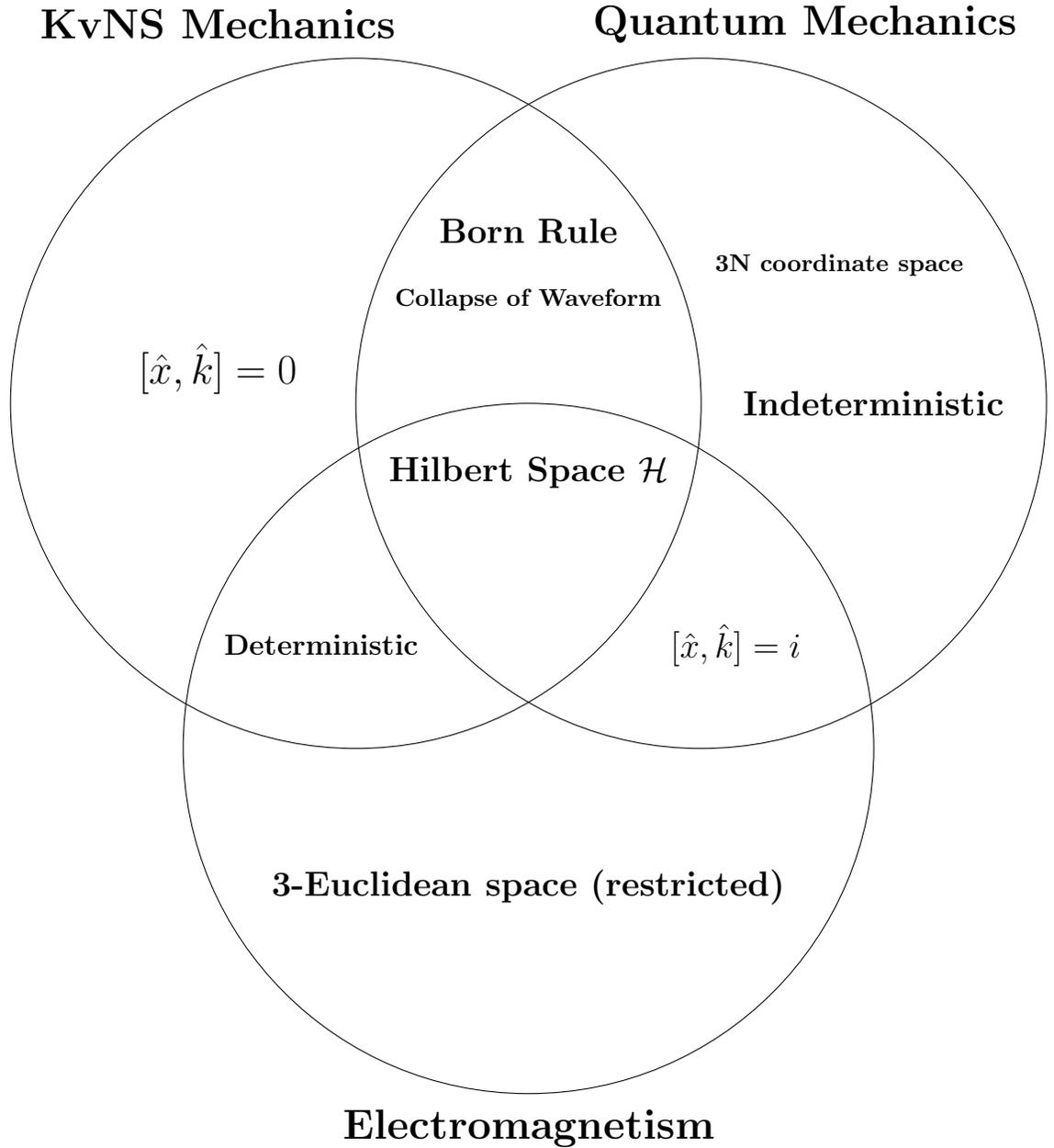

\subsection{Further identities and orthogonal expansions}
Based on the axioms of the theory, we introduce some further concepts and notation. We will give continuous vectors a continuous orthonormal basis, just like in both classical and quantum theories.
The classical Hilbert Space would contain:

\begin{equation}\label{eq:chi}
    \braket{\zeta'|\zeta} = \delta(\zeta'-\zeta) ~~\begin{cases} 
           ~\braket{x'|x} = \delta(x'-x) \\
          ~\braket{k'|k} = \delta(k'-k) 
       \end{cases}
\end{equation}
where $\zeta$ is a continuous variable, $x$ represents position, and $k$ is the wavenumber of electromagnetism. Identifying an orthonormal, denumberable (instead of continuous) basis set $\{\ket{U_n}\}$ denoted with the Kronecker delta:

\begin{equation}\label{eq:ortho}
    \braket{U_n|U_m} = \delta_{nm}
\end{equation}
We can also identify closure for the classical Hilbert Space, for both continuous and discrete states:

\begin{equation}\label{eq:closure}
    \int ~d\zeta \ket{\zeta}\bra{\zeta} = \hat{\mathbb{I}}~~\begin{cases} 
           ~\int ~dx \ket{x}\bra{x} = \hat{\mathbb{I}} \\
          ~\int ~dk \ket{k}\bra{k} = \hat{\mathbb{I}}
       \end{cases}
\end{equation}

\begin{equation}
\sum_n \ket{U_n}\bra{U_n} = \hat{\mathbb{I}}   
\end{equation}
The same expressions exist for both quantum and KvNS classical mechanics, as these are all Hilbert Space theories obeying the spectral theorem \citep{byron_and_fuller,hilbertoperators,gallone}. 

From these simple quantum-like relationships follow many very commonly known electromagnetic relationships. A common practice in a graduate electromagnetism course is to take the potential, for instance, and expand it along an orthogonal set of functions \citep[section 2.8]{jackson}. Any square integratable potential $\ket{A_\mu}$ is expandable in the following fashion:

\begin{equation}\label{eq:expansion}
\braket{\zeta|A_\mu}=\sum_n \braket{\zeta|U_n}\braket{U_n|A_\mu}
\end{equation}\hfill \citep[eq. 2.33]{jackson}\newline\newline
where $\braket{\zeta|U_n}$ represents the orthogonal functions of continuous variable $\zeta$ and $a_n = \braket{U_n|A_\mu}$ are the coefficients for each term, given by:

\begin{equation}\label{eq:coeff}
    a_n = \int d\zeta \braket{U_n|\zeta}\braket{\zeta | A_\mu}
\end{equation}\hfill \citep[eq. 2.32]{jackson}\newline\newline
Orthogonality of $\braket{\zeta|U_n}$ is easily demonstrated from Eq. \ref{eq:ortho}:
$$ \int d\zeta ~U_n^*(\zeta)U_m(\zeta)  = \delta_{nm}$$\hfill\citep[eq. 2.29]{jackson}
$$\sum_n \braket{\zeta'|U_n}\braket{U_n|\zeta} = \delta(\zeta'-\zeta)$$ \hfill \citep[eq. 2.35]{jackson}\newline\newline
Like in quantum mechanics, $\braket{\zeta|U_n}$ is defined in electromagnetism through an eigenvalue problem, i.e., through the Sturm–Liouville theory (see Appendix A and \citealt{Hanson_and_Yakovlev}).

This generalizes well to the multivariable case, where like in standard quantum mechanics we use the tensor product to bind together different vector spaces (third postulate from electromagnetic axioms), and from that the same familiar electromagnetic relations emerge (demonstrated for continuous variables $\zeta$ and $\eta$):

$$\ket{\boldsymbol{x}} = \ket{x_1, x_2,..., x_N} \equiv \ket{x_1}\otimes\ket{x_2}\otimes...\otimes\ket{x_N}$$\newline
$$A_\mu(\zeta,\eta) = \braket{\zeta,\eta|A_\mu} = \sum_{mn}\braket{\zeta|U_n}\braket{\eta|V_m}\braket{U_n,V_m|A_\mu}$$\hfill \citep[eq. 2.38]{jackson}\newline
where 
$$a_{mn} = \braket{U_n,V_m|A_\mu} = \int d\zeta \int d\eta ~U_n^*(\zeta)V_m^*(\eta) A_\mu(\zeta,\eta)$$\hfill\citep[eq. 2.39]{jackson}\newline

An example of an important orthogonal function for electromagnetism \citet[pp. 67]{jackson} discusses is the complex exponential:

\begin{equation}\label{eq: important Ux}
    \braket{x|U_n} = \frac{1}{\sqrt{a}}e^{i(2\pi nx/a)}
\end{equation}\hfill (\citealt{jackson} eq. 2.40)\newline\newline
where $n = 0, \pm 1, \pm 2, ...\in \mathbb{Z}$, is defined on interval $(-a/2, a/2)$. \citet[pp.67]{jackson} discusses how taking the limit of $a$ goes to infinity causes the set of orthogonal functions $\braket{x|U_n}$ to transform into a set of continuous functions. Essentially, the denumerable kets $\ket{U_n}$ transform to a continuum ket $\ket{k}$, related to the classical wavenumber. We would have:

\begin{equation}\label{eq:transf}
    \begin{cases} 
           \ket{U_n} \rightarrow \ket{k} \\
          \frac{2\pi n}{a} \rightarrow k \\
          \sum_n \rightarrow \int_{-\infty}^{\infty} dn = \frac{a}{2\pi}\int_{-\infty}^{\infty} dk \\
          a_n \rightarrow \sqrt{\frac{2\pi}{a}} a(k)
       \end{cases}
\end{equation}\hfill (Based on \citealt{jackson}, eq. 2.43)\newline\newline
The Kronecker delta will be replaced with the Dirac delta functional in all relevant expressions. Using eqs. \ref{eq:expansion} and \ref{eq:coeff} with \ref{eq:transf} gives us the famous Fourier Integral:

\begin{equation}\label{eq:Fourier}
    \braket{x|A_\mu} = \frac{1}{\sqrt{2\pi}}\int dk ~a(k)~ e^{ikx} = \int dk \braket{x|k}\braket{k|A_\mu}
\end{equation}\hfill\citep[eq. 2.44]{jackson}\newline\newline
with 
\begin{equation}\label{eq:Fourier_2}
    a(k) = \braket{k|A_\mu} = \frac{1}{\sqrt{2\pi}}\int dx~ e^{-ikx}~A_\mu (x) = \int dx \braket{k|x}\braket{x|A_\mu}
\end{equation}\hfill\citep[eq. 2.45]{jackson}\newline\newline
We have utilized closure (eq. \ref{eq:closure}) for the above two expressions. From eqs. \ref{eq:Fourier} and \ref{eq:Fourier_2}, we can produce an expression for $\braket{x|k}$, which implies a surprising quantum-like wavenumber position commutator for a purely classical wave. The interpretation of this will be explored in the next section. 

Another important orthogonal function is the spherical harmonics. The spherical harmonics can be identically computed and used in both the classical Hilbert Space and the quantum Hilbert Space. They play a central role in both fields, but are used to different ends. 
The identical quantum/classical Hilbert relations for $Y^m_l(\theta,\phi)$ are as follows:
$$\braket{\theta,\phi|lm} \coloneq Y_l^m(\theta,\phi)$$
$$\sum \ket{lm}\bra{lm}=\hat{\mathbb{I}}, \braket{l'm'|lm} = \delta_{mm'}\delta_{ll'} $$
$$\int d\Omega \ket{\theta,\phi}\bra{\theta,\phi} = \hat{\mathbb{I}}$$
The spherical harmonics for classical potentials also obey the usual

$$\bra{\theta,\phi}\hat{L}^2\ket{lm}=\Big[-\frac{1}{\sin^2\theta}\frac{\partial}{\partial \theta}\Big(\sin\theta\frac{\partial}{\partial \theta}\Big) -\frac{1}{\sin^2\theta}\frac{\partial^2}{\partial \phi^2}\Big]\braket{\theta,\phi|lm}=l(l+1)\braket{\theta,\phi|lm}$$
$$\bra{\theta,\phi}\hat{L}_z\ket{lm} = -i\frac{\partial}{\partial \phi}\braket{\theta,\phi|lm}=m\braket{\theta,\phi|lm}$$
These eigenvalue problems are, in fact, how the spherical harmonics are defined.
The spherical harmonics form a complete set of orthogonal functions, so they can be used as a basis for expansion. For a configuration of spherical symmetry, a typical expansion for the electric potential (in a region with no charge singularities) is:

$$V(r, \theta, \phi) = \braket{r, \theta, \phi| A_0} = \sum_{l=0}^n\sum_{m=-l}^{m=l} \braket{\theta, \phi |lm}\bra{lm}\otimes\braket{r|A_0} = \sum_{l=0}^n\sum_{m=-l}^{m=l} V_{lm}(r)Y_l^m(\theta, \phi) $$
where our weighted prefactor $V_{lm}$ is

$$V_{lm}(r) = \int d\Omega \braket{lm|\theta, \phi}\bra{\theta, \phi}\otimes\braket{r|A_0} = \int d\Omega ~Y_l^{m*}(\theta,\phi)V(r,\theta,\phi)$$
In the quantum Hilbert Space, $\braket{\theta,\phi|l,m}$ famously takes central stage in understanding the spectrum of the Hydrogen atom \citep{Shankar, Sakurai}. 
%
%
%
Other eigenfunctions of note useful for Hilbert Space electromagnetism can be found summarized in Appendix A.

\section{A Classical Position-Wavenumber Commutator and Feynman's Proof of Maxwell's Equations}

In both quantum mechanics (for the quantum waves) and KvNS classical mechanics (for classical point particles), we start with the position and momentum eigenvalue expressions as postulates for our theory \citep{dirac, Shankar, ODM, Sakurai, PiaseckiThesis}.

$$\hat{x}\ket{x} = x\ket{x}$$
$$\hat{p}\ket{p} = p\ket{p} \leftrightarrow \hat{k}\ket{k} = k\ket{k}$$
Here, for this new Hilbert Space theory for electromagnetism, we start in the same way, paralleling the \textit{classical} KvNS   complex Hilbert Space theory. Eqs. \ref{eq:Fourier} and \ref{eq:Fourier_2} are pulled directly from \citet{jackson}, as shown in the previous section. From Jackson's exposition, we can see immediately that there exists a wavenumber-position inner product in one dimension, given by

\begin{equation}\label{eq: xk}
    \braket{x|k} = \frac{1}{\sqrt{2\pi}}e^{ikx}.
\end{equation}
We provide two derivations that for the position and wavenumber eigenvalue problems, this directly leads to the commutation relationship $[\hat{x},\hat{k}]=i$.

\subsection{Derivation I: Utilization of properties of Dirac delta functional}

We first establish that

\begin{equation}\label{eq:correspondence}
[\hat{x},\hat{k}]=i \leftrightarrow \braket{x|k} = \frac{1}{\sqrt{2\pi}}e^{ikx}
\end{equation}
To do, we begin with
\begin{equation}\label{eq:start}
    \bra{x'}[\hat{x},\hat{k}]\ket{x}=(x'-x)\bra{x'}\hat{k}\ket{x}=i\delta(x'-x)
\end{equation}
We utilize the fact that a distribution $A$ that is zero everywhere except at one point $x_0$ can be expanded in terms of derivatives of Dirac delta functionals (for example, see \citealt{generalized_functions}), similar in form to a Taylor expansion: 

\begin{equation}\label{eq:theorem}
    A(x) = \sum_{n=0}^\infty a_n \delta^{(n)}(x-x_0)
\end{equation}
With the identity $x\delta'(x)=-\delta(x)$, we obtain

$$\bra{x'}\hat{k}\ket{x} = -i\delta'(x'-x)$$
\begin{equation}\label{eq:kxk}
    k\braket{x|k} = \bra{x}\hat{k}\ket{k} = \int dx' ~\bra{x}\hat{k}\ket{x'}\braket{x'|k} = -i\int dx' ~ \delta'(x'-x)\braket{x'|k}
\end{equation}
The definition of the rightmost integral above gives us:
\begin{equation}\label{eq:operator}
    k\braket{x|k}=-i\frac{\partial}{\partial x}\braket{x|k}
\end{equation}
This, of course, has the solution 

$$\braket{x|k}=Ce^{ikx}$$
where $C$ is the constant of integration. To identify the constant, we continue with eq. \ref{eq:chi} and the definition of the Dirac functional:

$$\delta(x'-x) = \braket{x'|x} = \int dk ~ \braket{x'|k}\braket{k|x} = C^2(2\pi)\delta(x'-x)$$
Ergo, we recover for one dimension

$$\braket{x|k} = \frac{1}{\sqrt{2\pi}}e^{ikx}$$
We went from $[\hat{x}, \hat{k}] = i$ to the Fourier expression for $\braket{x|k}$, but could have just as easily went in reverse. The result is completely general. 

\subsection{Derivation II: Wave Operator $\hat{k}$ as Generator of Electromagnetic Motion}

Starting from the commutator $[\hat{x},\hat{k}]=i$, we begin by constructing an anti-adjoint operator, defined by:

$$\hat{K} \coloneq -i\hat{k} $$
Since $[\hat{x},\hat{K}]=\hat{\mathbb{I}}$, this implies

$$[\hat{x},e^{\delta x \hat{K}}] = \delta x \cdot e^{\delta x \hat{K}}$$
Acting on a position ket $\ket{x}$, the above defined operator $e^{\delta x \hat{K}}$ would displace the state from $\ket{x}$ to $\ket{x + \delta x}$ as a generator of wave motion.

\begin{equation}\label{eq:generator}
    e^{\delta x \hat{K}}\ket{x} = \ket{x + \delta x}
\end{equation}
%
%
Using the expression \ref{eq:generator}, it is a simple task to Taylor expand each side:

$$\hat{\mathbb{I}}\ket{x} + \delta x\hat{K}\ket{x} + \mathcal{O}(\delta x^2)\ket{x} = \hat{\mathbb{I}}\ket{x}+ \delta x\frac{\partial}{\partial x}\ket{x} + \mathcal{O}(\delta x^2)\ket{x}$$
Comparing like terms and sandwiching from the left with a wavenumber bra $\bra{k}$, it is trivial that:

$$\hat{K}\braket{k|x} = -i\hat{k}\braket{k|x} = \frac{\partial}{\partial x}\braket{k|x}$$

\begin{equation}\label{eq: momentum wave operator}
    k\braket{x|k}=-i\frac{\partial}{\partial x}\braket{x|k}
\end{equation}
From here, the derivation proceeds exactly as in the previous section, with the same normalization used to derive expression \ref{eq: xk}. Quantum textbooks such as Sakurai use the same sequence of steps to derive the standard quantum generators of motion \citep[pp.40-64, 152-175, etc.]{Sakurai}. The wave commutator $[\hat{x},\hat{k}] = i$ therefore automatically implies expression \ref{eq: xk} (and for a Hilbert Space of operators and commutators, vice versa). 


%

\subsection{Feynman's Derivation of Maxwell's equations}
Feynman gave an interesting proof of Maxwell's equations based on the quantum position-momentum commutator and Newton's second law \citep{Dyson}. This strange usage of a quantum mathematical object for a classical field is not strange from the perspective of this work. The presence of this commutator in this purely classical Hilbert Space is a consequence of wave behavior (eq. \ref{eq:correspondence}), and Feynman applying the commutator to Newton's second law is arguably imposing wave behavior on Newtonian mechanics. Maxwell's equations as a consequence reflect the internal self-consistency of physical law, manifesting as the electric and magnetic force fields \citep{Dyson}. 

As \citet{Dyson} points out, the commutator relationship alternatively implies the existence of a vector potential $\boldsymbol{A}$ which obeys

$$[\hat{x}_j, \hat{A}_k] = 0$$
Even though Feynman derives Maxwell's Laws (which depend on the concept of the force field), it is very parsimonious with the $A_\mu = (V, \boldsymbol{A})$ approach we adopt here, as a more fundamental aspect of nature, and has been argued by many \citep[etc.]{ABeffect,AB1961,ABFeynman,Calkin1,Konopinski,mead}. 

The most unusual aspect of Feynman's derivation is the fact that it starts with a nonrelativistic version of Newton's second law and arrives at relativistic Maxwell equations \citep{Dyson}. A fully relativistic version of Feynman's proof is presented in Appendix B, using the language of quantum/classical Lie commutators.

\citet{OpenDirac,relODM} were able to utilize ODM and the KvNS formalism to derive the relativistic Dirac equation and classical Spohn equation for spin 1/2 particles based on the Hilbert Space. In their work, they used the canonical commutator to derive the Dirac equation, with the only difference being that the momentum was now the relativistic momentum. 

\section{Hermitian Operator and State Ket Representations of Four-Potential Elements}

In quantum mechanics, one sees the electromagnetic potentials represented as Hermitian operators $\hat{V}$ and $\boldsymbol{\hat{A}}$ acting on position kets. One may naturally wonder if there is a relationship between these operators and statements presented in this paper, such as those in eqs. \ref{eq:closureV} and \ref{eq:closureA}. There is, in fact, a very natural way one can interpolate between the operatorial forms and bra-ket representations. 

One can quite straightforwardly demonstrate that the two representations are related by

\begin{equation}
    \ket{A_0} = \int d\boldsymbol{x} ~\hat{V}\ket{\boldsymbol{x}}
\end{equation}
\begin{equation}
    \ket{\boldsymbol{A}}=\int d\boldsymbol{x} ~ \boldsymbol{\hat{A}}\ket{\boldsymbol{x}}
\end{equation}
or, equivalently, 

\begin{equation}
    \hat{V}=\ket{A_0}\bra{\boldsymbol{x}}
\end{equation}
\begin{equation}
\boldsymbol{\hat{A}}=\ket{\boldsymbol{A}}\bra{\boldsymbol{x}}
\end{equation}
This signifies that $\ket{A_0}$, for example, is constructed out of a continuous sum of scaled weighted positions $\ket{x}$, as one might expect.
This construction guarantees that the inner product for the potentials (eqs. \ref{eq:closureV} and \ref{eq:closureA}) will always be real-valued. 


\section{Fourier Analysis of the Four-Potential Elements}

We can generalize the one dimensional $\braket{x|k}$ into multiple dimensions using the tensor product familiar to quantum mechanics:

\begin{equation}\label{eq:3d xk}
    \braket{\boldsymbol{x}|\boldsymbol{k}} = \braket{xyz|k_xk_yk_z}=\frac{1}{(2\pi)^{3/2}}e^{i\boldsymbol{k}\cdot\boldsymbol{x}}
\end{equation}
This Hilbert Space represents a 3-Euclidean space, as is also apparent in Feynman's derivation of Maxwell's equations using the commutator (where $i,j,k = 1,2,3$).

Two common Fourier analysis cases in electromagnetism are when we restrict the potentials to a finite volume vs analyze the potentials over all spaces. In the case of a restricted volume of interest for the vector potential, it is expanded in the discrete basis $\ket{U_mV_nW_p}$:

$$\boldsymbol{A}(\boldsymbol{x},t) = \sum_{mnp} \braket{x|U_m}\braket{y|V_n}\braket{z|W_p}\braket{U_mV_nW_p|\boldsymbol{A}(t)} $$
where $\braket{\boldsymbol{x}|U_mV_nW_p}$ would represent the three-dimensional version of equation \ref{eq: important Ux} with 
$$\boldsymbol{k} =\frac{2\pi}{a}[m ~~n~~ p]$$
where $m, n, p = 0, \pm 1, \pm 2,...$. The standard coefficients of this expansion are given by $\boldsymbol{a}_{\boldsymbol{k}}(t) = \braket{U_mV_nW_p|\boldsymbol{A}(t)}$. This is what we expect for the magnetic potential \citep[eq. 10.2-9 and pp. 467-468]{vecAexp}. Likewise, for the scalar potential, we expect similarly

$$V(\boldsymbol{x},t) = \sum_{mnp} \braket{x|U_m}\braket{y|V_n}\braket{z|W_p}\braket{U_mV_nW_p|A_0(t)}, $$
which is the correct result \citep[sections 2.8-2.9,etc]{jackson}.

In the continuum limit, we once more switch the $\ket{U_mV_nW_p}$ with $\ket{k_xk_yk_z}$ (eq. \ref{eq:transf} but in three dimensions). The Hilbert Space representation provided here gives a simple and easy method to derive all Fourier expressions by simple expansions utilized in quantum calculations:

$$\braket{\boldsymbol{x}|A_\mu} = \int d^3k \braket{\boldsymbol{x}|\boldsymbol{k}}\braket{\boldsymbol{k}|A_\mu}$$
with coefficients 
$$\braket{\boldsymbol{k}|A_\mu} = \int d^3x \braket{\boldsymbol{k}|\boldsymbol{x}}\braket{\boldsymbol{x}|A_\mu}$$
where equation \ref{eq:3d xk} may be utilized.

\section{Treatment of Green's Operator}

The Hilbert Space formalism of electromagnetism lets us utilize Green's functions in the same fashion as in quantum mechanics (see also lengthy discussion in \citealt{Hanson_and_Yakovlev}). For the time-independent Green's function, we begin with the same standard definitions \citep[section 1.1]{economou}:

\begin{equation}\label{eq:Gdef}
    [z - \hat{L}]\hat{G}(z) = \hat{\mathbb{I}}
\end{equation}\hfill \citep[eq. 1.1]{economou}\newline\newline
where $\hat{G}$ is the Green's operator, $z \in  \mathbb{C}$ exists as a parameter, and $\hat{L}$ is a Hermitian operator with a complete set of eigenkets $\{\ket{L}\}$ obeying the eigenvalue problem

\begin{equation}
    \hat{L}\ket{L}=L\ket{L}
\end{equation}\hfill \citep[eq. 1.2]{economou}\newline\newline
We note that this can be either a classical or quantum eigenvalue problem (see \citealt{Hanson_and_Yakovlev}). 
In general, the form of $\hat{L}$ follows from the differential problem we are trying to solve \citep{economou}. 

The operators are defined in such a  manner that

\begin{equation}
    \bra{\boldsymbol{x}}\hat{G}(z)\ket{\boldsymbol{x'}} \equiv G(\boldsymbol{x}, \boldsymbol{x'}; z) 
\end{equation}
\begin{equation}
    \bra{\boldsymbol{x}}\hat{L}\ket{\boldsymbol{x'}}\equiv \delta(\boldsymbol{x}-\boldsymbol{x'})L(\boldsymbol{x})
\end{equation}
resulting in the usual Green's function expression:

$$[z-L(\boldsymbol{x})]G(\boldsymbol{x},\boldsymbol{x'};z) = \delta({\boldsymbol{x}-\boldsymbol{x'}})$$
As a consequence of the spectral theorem, we also know that

\begin{equation}
    \braket{L|L'} = \delta_{LL'}
\end{equation}\hfill (\citealt[eq. 1.3]{economou}; \citealt[eq. 3.155]{jackson})
\begin{equation}\label{eq:closureL}
    \sum \ket{L}\bra{L} = \hat{\mathbb{I}}
\end{equation}\hfill (\citealt[eq. 1.3]{economou}; utilized in \citealt[eqs. 3.157,8]{jackson})\newline\newline
We can begin by solving eq. \ref{eq:Gdef} to get the expression for the Green's operator

$$\hat{G}(z) = \frac{\hat{\mathbb{I}}}{z-\hat{L}}$$
With eq. \ref{eq:closureL}, it can be quickly seen that (in a continuous or denumberable basis) one can write

\begin{equation}
    \hat{G}(z) = \sum_n \frac{\ket{L}\bra{L}}{z-L}+\int dc~ \frac{\ket{L_c}\bra{L_c}}{z-L_c}
\end{equation}\hfill \cite[eq. 1.11]{economou}\newline\newline
where the subscript $c$ represents continuous states. 

With the position basis, one finds the standard electromagnetic expansions of the Green's function

\begin{equation}\label{eq:greensmoo}
    G(\boldsymbol{x}, \boldsymbol{x'}; z) = -4\pi \sum_n \frac{\psi^*_n(\boldsymbol{x'})\psi_n(\boldsymbol{x})}{z-L} -4\pi \int dc \frac{\psi^*_{nc}(\boldsymbol{x'})\psi_{nc}(\boldsymbol{x})}{z-L_c} 
\end{equation}\hfill (\cite{jackson} eq. 3.160)\newline\newline
where $\psi_n(\boldsymbol{x}) = \braket{\boldsymbol{x}|L}$ and  $\psi_{nc}(\boldsymbol{x}) = \braket{\boldsymbol{x}|L_c}$ (prefactor of $4\pi$ is conventional for cgs units). One can therefore equivalently represent the Green's function in terms of orthonormal eigenkets and eigenvalues in both electromagentism \citep{jackson,Hanson_and_Yakovlev} and quantum mechanics \citep{economou}.

An example of a commonly used differential is the Laplace operator for the electrostatic potential:

$$\boldsymbol{\nabla}^2V=-4\pi \rho_e$$
(in cgs units). With $L(\boldsymbol{x}) = - \boldsymbol{\nabla}^2$ over all 3-Euclidean space, we identify $\ket{L}$ with $\ket{k}$ (\citealt{economou} pp. 9; also see eq. 1.34) from the fact that

$$-\boldsymbol{\nabla}^2\bra{\boldsymbol{x}}=\bra{\boldsymbol{x}}\boldsymbol{\hat{k}}^2$$
Then:

\begin{equation}\label{eq:Gint}
    G(\boldsymbol{x}, \boldsymbol{x'};z)=\int \frac{d\boldsymbol{k}}{(2\pi)^3}\frac{\braket{\boldsymbol{x|k}}\braket{\boldsymbol{k|x'}}}{z-k^2}
\end{equation}
Carrying out integrations and setting $z=0$ for the electric point source \citep{economou}, we retrieve as an illustration the well established fact from electrostatics

\begin{equation*}\label{eq:exofpot}
    V(\boldsymbol{x}) = -4\pi\int d\boldsymbol{x'}~G(\boldsymbol{x},\boldsymbol{x'};0)\rho_e(\boldsymbol{x'}) =\int d\boldsymbol{x}\frac{\rho_e(\boldsymbol{x})}{|\boldsymbol{x}-\boldsymbol{x'}|}
\end{equation*}\hfill (Based on \citealt{economou} eq. 1.44 and \citealt{jackson} eq. 3.164)\newline\newline
In the same fashion, we can compute from eq. \ref{eq:Gint} the frequently used Helmoltz equation Green's function (\citealt{economou} eq. 1.40):

\begin{equation*}
    G(\boldsymbol{x}, \boldsymbol{x}; k) = \frac{e^{\pm i\boldsymbol{k}\cdot
|\boldsymbol{x}-\boldsymbol{x'}|}}{4\pi|\boldsymbol{x}-\boldsymbol{x'}|}
\end{equation*}
%




%

%
The axiom of unitary time evolution can be used in the construction of time-dependent Green's operators for electromagnetism. It can be shown that the propagator equation (eq. \ref{eq:propogator}) is itself a Green's function with the above definition (i.e., see \citealt{economou} eq. 2.15 and description). The Green's function will be the solution of 

$$\Big(\frac{i}{c}\frac{\partial}{\partial t}-L(\boldsymbol{x})\Big)G(\boldsymbol{x},\boldsymbol{x'};t-t')=-4\pi\delta (\boldsymbol{x}-\boldsymbol{x'})\delta (t-t')$$
for first-order diffusion-type or Schr{\"o}dinger-type equations \citep{economou}. For second order (wave-like) equations, the Green's function will be the solution of \citep{economou}

$$\Big(-\frac{1}{c^2}\frac{\partial^2}{\partial t^2}-L(\boldsymbol{x})\Big)G(\boldsymbol{x},\boldsymbol{x'};t-t')=-4\pi\delta (\boldsymbol{x}-\boldsymbol{x'})\delta (t-t')$$
which leads to well-known solutions (using eq. \ref{eq:Gint} and time-frequency Fourier transform) of the Laplacian operator:

$$G(\boldsymbol{x},\boldsymbol{x'};t-t') = \frac{\delta(t'-(t-|\boldsymbol{x}-\boldsymbol{x'}|/c))}{|\boldsymbol{x}-\boldsymbol{x'}|}$$
Typical electromagnetic calculations can be carried out with Dirac notation in Hilbert Space. More on Green's operators for electromagnetism can be found in \citet{Hanson_and_Yakovlev}.

\section{Maxwell Fields in Hilbert Space}

So far in this paper, we have not introduced the important force description of electric $\boldsymbol{E}$ and magnetic fields $\boldsymbol{H}$. One can very simply define these fields as:
$$\ket{\boldsymbol{E}}=\ket{E}\ket{\boldsymbol{n}_e}$$
$$\ket{\boldsymbol{H}}=\ket{H}\ket{\boldsymbol{n}_h}$$
where adjacent to eq. \ref{eq:vector_potential}, we define $\ket{E}$,$\ket{H}$ as the scalar amplitude belonging to the $L^2$ space (eq. \ref{eq:space_L}), and $\ket{\boldsymbol{n}_h}$,$\ket{\boldsymbol{n}_e}$ are the Euclidean portions living in $\mathbb{R}^3$. These are just well-known transformations of the gauge fields (cgs):
$$\ket{\boldsymbol{E}(t)} = -\frac{i}{c}\hat{\Omega}\ket{\boldsymbol{A}(t)} - i\boldsymbol{\hat{k}}\ket{A_0(t)}$$
$$\ket{\boldsymbol{H}}=\hat{\Gamma}\ket{\boldsymbol{A}}$$
$$\braket{e_l|\boldsymbol{H}}=\bra{e_l}\hat{\Gamma}\ket{e_k}\braket{e_k|\boldsymbol{A}}$$
with Einstein summation ($\ket{e_i} \in \mathbb{R}^3$), with the curl matrix defined as
$$\hat{\Gamma} := \begin{bmatrix}
    0 & i\hat{k}_z & -i\hat{k}_y \\
    -i\hat{k}_z & 0 & i\hat{k}_x \\
    i\hat{k}_y & -i\hat{k}_x & 0
\end{bmatrix}$$
%
The above definitions of electric and magnetic force fields are gauge-invariant \citep{jackson}. Orthonormal expansions can be then carried out for electromagnetic fields, not just potentials, in similar manner thanks to the above expressions \citep[]{jackson}. It is impotant to stress that although a great deal of attention has been paid to the potentials, the field expansions are a natural consequence of the concepts discussed up to this point. They too have a very natural home in this Hilbert Space theory.




\section{Comparison of Quantum and Classical Wave Hilbert Space Theories}

In the absence of current and charges, the four-potential propagates as a wave equation: 

$$\boldsymbol{\nabla}^2 A_\mu - \frac{1}{c^2}\frac{\partial^2 A_\mu}{\partial t^2} = 0$$
Because of this, it is not surprising that we can propose a Hilbert Space with wave behavior for $A_\mu$. Just like the wavefunction, for example, the point charge electric potential $V$ and its derivative go to zero at infinity. Through QED, we know that the $1/r$ law for point potential breaks down at small values of $r$, avoiding the singularity encountered at $r=0$. It is not, therefore, surprising that elements of the four-potential can be square normalizable in general and therefore are elements of a Hilbert Space \citep{byron_and_fuller,hilbertoperators,gallone}. In the laboratory, there are no infinities. 

Unlike the Hilbert Space in quantum mechanics, we do not impose any probabilistic interpretation on the mathematics. Electromagnetism is still presented as a deterministic theory. Therefore, a large difference between the classical wave and quantum theory is that the quantum theory has a probabilistic interpretation, but the classical wave theory is deterministic due to lack of such an imposition. The mathematical structure, however, is identical. 

A purely classical commutator relationship for wavenumber (or momentum) and position operators was derived for the electromagnetic Hilbert Space (section 3). The derivations for eq. \ref{eq:correspondence} highlight the fact that commutation structure implies a wavenature for the two observables (and vice versa). Since electromagnetism is a wave theory (albeit classical), this is not surprising on a mathematical level. The existence of the classical commutator structure further highlights that particle-wave uncertainty in quantum mechanics emerges from the wave-like nature of the quantum. The commutator for electromagnetism and the canonical commutator are also identical in shape. Since de Broglie has shown that 
$\boldsymbol{p} = \hbar \boldsymbol{k}$, it is not difficult to see that one goes from the electromagnetic commutator $[\hat{x},\hat{k}]=i$ to the quantum canonical commutator $[\hat{x},\hat{p}]=i\hbar$ by simply multiplying both sides by $\hbar$. The difference between classical and quantum mechanics is not in the existence between a commutator of position and wavenumber, but in the presence of the reduced Planck unit $\hbar$, i.e., in the existence of the Planck-Einstein and de Broglie relations.

One might therefore pen a classical analogue to the Heisenberg Uncertainty Principle. From introductory quantum mechanics \citep{Shankar, Sakurai}, it is possible to show that for any two Hermitian operators $\hat{A}$ and $\hat{B}$, the uncertainty principle would be \citep{Sakurai}:

\begin{equation}\label{eq:uncertainty}
    \sigma_A\sigma_B \geq \frac{|\bra{\psi}[\hat{A},\hat{B}]\ket{\psi}|}{2}
\end{equation}
%
%
This, of course, assumes that the classical (KvNS) or quantum $\ket{\psi}$ is related to a probability distribution, since we must define an expectation value of observables \citep{dirac,von_N_textbook}. For classical fields $\ket{A_0}$ and $\ket{\boldsymbol{A}}$, there is no Born Rule analogue for traditional electromagnetism. Even though both quantities are square integrable in this scheme, there is no known physical significance ascribed to, for example, $\braket{A_0|\boldsymbol{x}}\braket{\boldsymbol{x}|A_0}$ (analogous to $\psi^*\psi$). Since the gauge fields are also non-unique without choice of gauge, it seems very unlikely that the amplitude squared can be mapped into a classical probability density. Hence, this is why a Born Rule for the classical gauge fields is problematic, and excluded from the initial list of axioms for $\ket{A_0}$ and $\ket{\boldsymbol{A}}$.

However, we may draw a \textit{correspondence} between the physical feature of electromagnetic energy density and the spread of the wave \citep{classicaluncertaintyprinciple}. Using section 7, we can write the electromagnetic wave to be (cgs):

\begin{equation}\label{eq:EM_wave}
    \ket{\Psi} = \frac{1}{\sqrt{8\pi}}\ket{E}\ket{\boldsymbol{n}_{||}}+\frac{1}{\sqrt{8\pi}}\ket{H}\ket{\boldsymbol{n}_{\perp}}
\end{equation}
where $\ket{E}$ and $\ket{H}$ are abstract representations of the scalar amplitudes of the $\boldsymbol{E}$ and $\boldsymbol{H}$ fields. The energy density becomes:

$$\rho_{\epsilon} = \braket{\Psi|\boldsymbol{x}}\braket{\boldsymbol{x}|\Psi}$$
which allows us to write quantum-like wave expectation values for classical observables:
\begin{equation*}
    \braket{\zeta} = \frac{\bra{\Psi}\hat{\zeta}\ket{\Psi}}{\braket{\Psi|\Psi}} = \frac{\int d\zeta ~ \zeta~ \rho_\epsilon(\zeta)}{\int d\zeta ~  \rho_\epsilon(\zeta) }
\end{equation*}
Armed with this correspondence, the physical spread (variance) of the wave properties can be written as

$$\sigma_\zeta = \bra{\Psi}\hat{\zeta}^2\ket{\Psi}-\bra{\Psi}\hat{\zeta}\ket{\Psi}^2$$
Utilizing the Cauchy–Bunyakovsky–Schwarz inequality (as shown in the uncertainty derivation in \citealt{PiaseckiThesis}), one once again arrives at eq. \ref{eq:uncertainty}, but from a purely classical standpoint. (It is important to stress here that there is no probability interpretation  utilized for the classical field; we use these distributions to measure the \textit{physical spread} of the waves in actuality, not predict the probabilities for features to occur in repeated experiments. There is no collapse postulate.)

This analogue to the Heisenberg Uncertainty Principle has actually been known for some time, but not widely recognized as such. For instance, \cite{jackson} demonstrates that 
amplitude normalizable electromagnetic waves have the property

\begin{equation}\label{eq:EMHUP}
    \sigma_x \sigma_k \geq \frac{1}{2},
\end{equation}\hfill \citep[eq. 7.82]{jackson}

\hfill\cite[eq. 9.4.87]{classicaluncertaintyprinciple}\newline\newline
If we plug the electromagnetic commutator $[\hat{x},\hat{k}]=i$ into eq. \ref{eq:uncertainty}, we derive the exact same inequality.
A similar relationship exists for time and frequency of the electromagnetic wave \citep[pp. 301]{jackson}, mirroring the energy-time uncertainty principle of quantum mechanics. 
\citet{classicaluncertaintyprinciple} and \citet{Mansuripur} go into great detail how optical experiments bear out eq. \ref{eq:EMHUP} as a classical analogue to the uncertainty principle. 

This brings into sharp focus a common misunderstanding of the uncertainty principle. Many seem to mistakenly think that the principle captures some sort of indeterminism in nature, but really the indeterminism is found in the wave collapse postulate of quantum theory (breakdown in unitarity). What it captures is that for waves the momentum density of states product with the position density of states is fundamentally limited, unlike the case for classical point masses. Unlike localized point masses, waves have a position spread throughout space and will carry a spread of momenta. Here, the interpretation is the same. It can be seen that even in the classical case (eq. \ref{eq:EMHUP}) where we lack waves of probability we have a fundamental limit on the wavenumber times position density of states. Classical electromagnetism is still a purely deterministic theory.  


The KvNS formalism of classical mechanics has a commutator of form $[\hat{x},\hat{p}]=0$, since we are no longer dealing with waves or fields, but specifying infinitely precise point particles in the typical style of Newtonian mechanics. The ``fuzzy trajectories" of quantum theory are really just waves spread throughout space (as argued by \citealt{Hobson} and many others). The KvNS formalism allows for point particles by placing the wave uncertainty (spread) into other operators of the Koopman algebra - the ``unobservable" hidden variables for classical theory \citep{Sudarshan,ODM, McCauletal, PiaseckiThesis}. This then allows, for example, the derivation of the Liouville equation for point masses from the postulate of unitary time evolution in KvNS theory \citep{ODM}.
\begin{equation*}
    i\frac{\partial}{\partial t}\ket{\psi(t)} = \hat{L}\ket{\psi(t)} ~and~ [\hat{x},\hat{p}]=0 ~\Longrightarrow~ \frac{\partial}{\partial t}\rho(\mathbb{Q,P},t) = (\frac{\mathbb{P}}{m}\frac{\partial}{\partial \mathbb{Q}}-V'(\mathbb{Q})\frac{\partial}{\partial \mathbb{P}})\rho(\mathbb{Q,P},t)
\end{equation*}
where we use the classical probability density $\rho = \braket{\psi(t)|\mathbb{Q,P}}\braket{\mathbb{Q,P}|\psi(t)}$. The Liouville equation appears in the context of point masses.
Further cementing the wave verses point mass distinction, the amplitude and phase of the classical wavefunction in KvNS theory are completely separable, unlike the amplitude and phase in quantum theory and electromagnetism \citep{MauroPhDThesis}. 

In a similar vein, it is perhaps not too surprising that the same quantum-like operators are seen in both electromagnetism and quantum mechanics (equations \ref{eq:operator} and \ref{eq: momentum wave operator}). Schr{\"o}dinger's original inspiration for the form of the quantum operators was likely from electromagnetism. In his original papers proposing matter has a wave-like structure, de Broglie attempted to give electromagnetism and matter an equal footing in treatment through the lens of the then new Relativity theory \citep{Broglie_c,Broglie_a,Broglie_b, Broglie}. Schr{\"o}dinger, motivated by de Broglie's work \citep{Schrodinger}, penned his now famous equation, likely by deducing the wave operators by ansatz from the known classical wave equations. It turns out, in the electromagnetic Hilbert Space you also have identical $x$ and $k$ (or $p$) representations for your position and wavenumber (momentum) operators as you do in quantum mechanics. Starting from $[\hat{x},\hat{k}]=i$, we find in eqs. $\ref{eq:operator}$ and $\ref{eq: momentum wave operator}$ that we can write the position representation as

$$\hat{x}~ \dot{=}~x ~~~{and}~~~ \hat{k} ~\dot{=}~-i\frac{\partial}{\partial x}$$
However, if we began from the left side of eq. \ref{eq:start} with $\bra{k'}[\hat{x},\hat{k}]\ket{k}$ instead of $\bra{x'}[\hat{x},\hat{k}]\ket{x}$, we would have produced the wavenumber (momentum) representation of our operators:
$$\hat{x}~ \dot{=}~i\frac{\partial}{\partial k} ~~~{and}~~~ \hat{k} ~\dot{=}~k$$
The difference between the quantum and classical lies once again in the presence of $\hbar$.

These are the same mathematical objects in both electromagnetism, KvNS classical mechanics, and quantum mechanics. What differs is how we interpret and utilize the mathematical objects in electromagnetic verses quantum Hilbert Space. Both include superpositions of weighted orthonormal functions (eq. \ref{eq:expansion}), the usage of an eigenvalue problem to identify the orthogonal functions, an inner product between vectors in a dual space, Dirac delta and Kronecker delta representations of closure (in continuous and discrete bases), etc (see also \citealt{Hanson_and_Yakovlev}). Many mathematical objects, such as the spherical harmonics $\braket{\theta, \phi|lm}$, are identical in form in both classical and quantum spaces. 

The principle difference between the electromagnetic approach and the quantum approach is that the electromagnetic approach contains no probability density (e.g., a Born Rule for the gauge fields and wavefunction collapse), and therefore we do not utilize the Hilbert Space to make probabilistic predictions of the outcomes of systems, unlike in both quantum (for waves) and KvNS (for classical point masses). Collapse of the waveform is completely absent, unlike in both quantum and classical KvNS mechanics. 

Another difference between the three Hilbert Space theories is the spaces they represent. Quantum mechanics represents a 3N-coordinate space, KvNS is in 3N-phase space, and this electromagnetic theory represents a 3-Euclidean space. The wavefunction of quantum mechanics living in 3N-coordinate space has famously perplexed physicists such as Einstein and Schr{\"o}dinger. Einstein expressed frustration when he famously said ``Schr{\"o}dinger’s works are wonderful –
but even so one nevertheless hardly comes closer to a real understanding. The field in a many-dimensional coordinate space does not smell like something real" \citep{Howard}. Also: ``Schr{\"o}dinger is, in the beginning, very captivating. But the waves in n-dimensional coordinate space are indigestible..." \citep{Howard}. Schr{\"o}dinger, Lorentz, Heisenberg, Bohm, Bell, and others struggled with the same property of quantum fields \citep{Howard, Travis}. For the electromagnetic Hilbert Space, we do not face the same issues, since, for example, a scalar potential of the form of $\braket{x_1,x_2,...,x_{3N}|A_0}$ would still be interpreted as living in $\mathbb{R}^3$, even though it is a function on a configuration space (the $x_1,...,x_{3N}$ listed here are coordinates of the source charges that create the scalar field). The interpretation of the wavefunction $\braket{x_1,x_2,...,x_{3N}|\psi}$ is not so staightfoward, however, due to its probabilistic interpretation, and has been hotly debated for over a century \citep{Howard, Travis}. We will not shed any light on it here.


There are many ways the preliminary groundwork of this paper may be extended. It has been argued, for example, that there exist classical entanglement states in classical optics and that it is related to a Hilbert Space structure \citep{Spreeuw,CE14,CE14b,Qian,CE16}. This brings up a very interesting question how Bell's Inequality relates to classical optics \citep{CE14,CE14b,Qian,CE16}. This potential-based formulation might also extend to traditionally quantum-specific algorithms, as recently proposed for KvNS mechanics \citep{simkvn}. It would be interesting to connect this approach to classical scattering theory and the Optical Theorem. Another intriguing question is if a close tie can be made between the Wigner function and its classical optical Wigner distribution. ODM has demonstrated \citep{wignerphasespace} that one can treat the Wigner function as a quantum probability amplitude projected into a point in \textit{classical} phase space. This raises the interpretation that the Wigner function's lack of positive definiteness is not problematic, since a probability amplitude can indeed take on negative values \citep{wignerphasespace}. The optical Wigner function may similarly emerge from an operator and commutator structure that captures both wave-like and ray (point-like) behavior - much like how the original Wigner function arises naturally in the context of KvNS point particle commutator (see \citealt{ODM,wignerphasespace}). The meaning of classical verses quantum behavior can be further explored as well in this operatorial framework. We plan on further developing these concepts in future works.

%

%
\section*{Acknowledgments}
I'd like to thank those who helped inspire this paper. Frank Tipler of Tulane University motivated this train of thought through discussions of the origin and meaning of the Uncertainty Principle. Denys Bondar of Tulane educated me in classical Hilbert Space theories; I am indebted to him for providing useful notes that helped me craft the proof in section 3.1. Also for providing papers of relevance to the project. David Aspnes of North Carolina State University went over many facets of classical electrodynamics with kindness and patience. Peter McGrath of NCSU discussed mathematics behind $L^2$ Hilbert Spaces. Chueng Ji of NCSU gave me the opportunity to present this material to his theory group and had many fruitful conversations on quantum material. Other fruitfull discussions with others and their feedback.

\section*{Appendix A: Brief Summary of Orthonormal Expansions for Hilbert Space Electromagnetism}

In section 2.2, the spherical harmonics were summarized for expansions in electromagnetism. Here, we cover some other useful functions for electromagnetic theory. These functions are defined through the standard Sturm–Liouville theory:

$$\Big[\frac{d}{dx}\Big(p(x)\frac{d}{dx}\Big)+q(x)\Big]\Lambda(x) = - \lambda r(x)\Lambda(x)$$
where $\lambda$ is the eigenvalue and $\Lambda$ is the eigenfunction. 
\newline\newline
\textbf{Bessel function of the first kind:}
\newline
Sturm–Liouville problem: 
$\frac{d}{d\rho}(\rho\frac{dJ_\nu}{d\rho})+(\rho-\frac{\nu^2}{\rho})J_\nu=0$
\newline
$J_\nu(x_{\nu n}\rho/a)= \braket{x_{\nu n}\rho/a|J_\nu}$ in Hilbert Space. \newline
Orthogonal function: $\braket{x_{\nu n}\rho/a|U_n} = \sqrt{\rho}J_\nu(x_{\nu n}\rho/a)$\newline
Orthogonality: $\braket{U_{\nu n}|U_{\nu m}}=\int^{a}_0d\rho~\rho J_\nu(x_{\nu n}\rho/a)J_\nu(x_{\nu m}\rho/a) = \frac{a^2}{2}[J_{\nu +1}(x_{\nu n})]^2\delta_{nm}$
\newline\newline
\textbf{Legendre Polynomial:}\newline
Sturm–Liouville problem: 
$\frac{d}{dx}((1-x^2)\frac{dP_\nu}{dx})+\nu(\nu-1)P_\nu=0$
\newline
$P_\nu(x) = \braket{x|P_\nu}$ in Hilbert Space. \newline
Orthonormal function: $\braket{x|U_\nu}= \sqrt{\frac{2\nu+1}{2}}\braket{x|P_\nu}$\newline
Orthogonality: 
$\braket{P_n|P_m}=\int^{1}_{-1}dx~P_n(x)P_m(x) = \frac{2}{2n+1}\delta_{nm}$
\newline
Resolution of identity:
$\hat{\mathbb{I}} = \sum_{n=0}^{\infty}\frac{2n+1}{2}\ket{P_n}\bra{P_n}$
\newline\newline
\textbf{Associated Legendre Polynomial:}\newline
Sturm–Liouville problem: 
$\frac{d}{dx}((1-x^2)\frac{dP_{\nu m}}{dx})+[\nu(\nu-1)-\frac{m^2}{1-x^2}]P_{\nu m}=0$
\newline
$P_{\nu m}(x) = \braket{x|P_{\nu m}}$ in Hilbert Space. 
\newline
Orthonormal function: $\braket{x|U_\nu} = \sqrt{\frac{2\nu+1}{2}\frac{(\nu - m)!}{(\nu+m)!}}\braket{x|P_{\nu m}}$
\newline
Orthogonality: 
$\braket{P_{\nu m}|P_{\nu' m}}=\int^{1}_{-1}dx~P_{\nu m}(x)P_{\nu' m}(x) = \frac{2}{2n+1}\frac{(\nu - m)!}{(\nu +m)!}\delta_{\nu\nu'}$
\newline
Resolution of identity:
$\hat{\mathbb{I}} = \sum_{m=-\nu}^{m=+\nu}\sum_{n=0}^{\infty}\frac{2n+1}{2}\frac{(\nu + m)!}{(\nu - m)!}\ket{P_{\nu m}}\bra{P_{\nu m}}$
\newline\newline
\textbf{Hermite Polynomial:}\newline
Sturm–Liouville problem: 
$\frac{d}{dx}(e^{-x^2}\frac{dH_\nu}{dx})+2\nu e^{-x^2}H_\nu=0$
\newline
$H_\nu(x) = \braket{x|H_\nu}$ in Hilbert Space. \newline
Orthogonal functions: $\braket{x|U_\nu} = \sqrt{e^{-x^2}}\braket{x|H_\nu}$\newline
Orthogonality: 
$\braket{U_n|U_m}=\int^{\infty}_{-\infty}dx~e^{-x^2}H_n(x)H_m(x) = \sqrt{\pi}2^n n! \delta_{nm}$
\newline
Resolution of identity: $\hat{\mathbb{I}} = \sum_{\nu=0}^{\infty}\frac{e^{-x^2}}{\sqrt{\pi}2^\nu (\nu!)}\ket{H_\nu}\bra{H_\nu} $

\section*{Appendix B: Relativistic Feynman's Proof of Maxwell's Equations}
Feynman's proof \citep{Dyson} captures an underlying physical structure, being able to reproduce the homogeneous Maxwell equations, although we believe it captures the physical structure imperfectly due to certain assumptions. This is the consensus view \citep[etc.]{feynmanmaxwel1990,feynmanmaxwel1992,feynmanmaxwell1993,feynmanmaxwel1995,feynmanmaxwel1996,feynmanmaxwel1999important,feynmanmaxwel2004,feynmanmaxwel2007,feynmanmaxwel2009,feynmanmaxwel2012}. It can be shown that it is possible to construct a relativistically consistent version of Feynman's proof, while at the same time avoiding its unsavory elements. 

Feynman's motivation was nobler than our own. Feynman was trying to establish a theory for the quantum based on the least number of assumptions as possible; we, however, are interested in the continuity of ideas between different branches of physics. We establish continuity, whereas Feynman was attempting (and ultimately failed in this regard) in building new physics. 

Almost all work following up on Feynman's curious proof relies on the Poisson bracket structure or Lagrange formalism. In the spirit of this Hilbert Space formalism (similar to the KvNS approach), we will do all classical calculations using quantum-like Lie commutators, like \citet{Dyson} used in the original paper. \citet{relODM} utilizes the commutator structure and mathematics of Hilbert Space to derive the Dirac equation. This approach also is utilized to apply the Dirac equation to open systems \citep{OpenDirac}. Following the original proof attributed to Feynman, we use classical commutators. 

Feynman's proof \citep{Dyson} begins with

\begin{equation}\label{eq:feynmannewtonssecondlaw}
    \dot{\hat{p}}_j = \hat{F}_j(\hat{x},\dot{\hat{x}},t)
\end{equation}
\begin{equation}\label{eq:feynmanpositionassumption}
    [\hat{x}_i,\hat{x}_j] = 0
\end{equation}
\begin{equation}\label{eq:cannonicalcommutatorfeynman}
    [\hat{x}_i,\hat{p}_j] = i\hbar\delta_{ij}
\end{equation}
and ends in 
\begin{equation}\label{eq:referto1}
    \hat{F}_j = \hat{E}_j + \epsilon_{jkl}\dot{\hat{x}}_k\hat{H}_l
\end{equation}
\begin{equation}\label{eq:referto2}
    div ~\hat{H} =0 
\end{equation}
\begin{equation}\label{eq:referto3}
    \frac{\partial\hat{H}}{\partial t}+ curl~\hat{E} = 0
\end{equation}
with the remaining two Maxwell equations left as definitions of charge density and current. Although unstated throughout the proof, Feyman sneaks in a fourth assumption (other than eqs. \ref{eq:feynmannewtonssecondlaw} - \ref{eq:cannonicalcommutatorfeynman}), and that is that the form of the momentum is 

\begin{equation}\label{eq:tacit}
    \hat{p}_j = m\dot{\hat{x}}_j,
\end{equation}
which is where the issue with the Feynman proof lies. This assumption is what leads to a contradiction. For one, it leads to the fact that the velocity components do not commute. Feynman defines the magnetic field to be:

\begin{equation}\label{eq:feymanmagfield}
    \hat{H}_l= -\frac{im^2}{2\hbar}\epsilon_{jkl}[\dot{\hat{x}}_j,\dot{\hat{x}}_k]
\end{equation}
However, a glaring issue with this is that it implies, with Feynman's tacit assumption (eq. \ref{eq:tacit}) , that $[\hat{p}_i,\hat{p}_j] \neq 0$, which contradicts a basic principle of quantum mechanics \citep[eq. 1.224]{Sakurai}. Although Feynman's proof adopts one quantum commutator (eq. \ref{eq:cannonicalcommutatorfeynman}), it neglects another commutator principle of quantum mechanics: $[\hat{p}_i,\hat{p}_j] = 0$. This can all be remedied, however, with a more appropriate definition of momentum. 

If instead we assume a relativistic momentum with minimal coupling \citep{feynmanmaxwel1999important}, we will still be able to carry out Feynman's proof in similar manner, achieving in the end the same conclusions (eqs. \ref{eq:referto1} - \ref{eq:referto3}), but free of contradiction and other unappealing features of the original proof. The price of this is not high, as we are just switching one unappealing assumption (eq. \ref{eq:tacit}) with a better one:

\begin{equation}\label{eq: mincoup}
    \hat{P}_j = {\hat{p}}_j + \hat{A}_j(\hat{x},t)
\end{equation}
where relativistic operators are assumed (like in \citealt{OpenDirac,relODM}).
At the same time, we also change eq. \ref{eq:feynmannewtonssecondlaw} to
\begin{equation}\label{eq:newtonssecondlaw}
    \dot{\hat{P}}_j = \hat{F}_j(\hat{x},\dot{\hat{x}},t)
\end{equation}
and eq. \ref{eq:cannonicalcommutatorfeynman} with
\begin{equation}\label{eq:cannonicalcommutatorfeynmanimproved}
    [\hat{x}_i,\hat{P}_j] = i\hbar\delta_{ij}
\end{equation}
\citet{feynmanmaxwel1999important} very convincingly argue that Feynman's proof is capturing minimal coupling behavior. This derivation will reflect that.

First, we will prove that the commutation of two components of eq. \ref{eq: mincoup} must be the magnetic field. Whereas Feynman left it as a simple definition (eq. \ref{eq:feymanmagfield}), by eq. \ref{eq: mincoup} it must necessarily follow. The commutation of the two gives us:

\begin{equation}
    [\hat{P}_i,\hat{P}_j] = [\hat{A}_i, \hat{p}_j]+[\hat{p}_i,\hat{A}_j] = -i\hbar\Big(\frac{\partial \hat{A}_i}{\partial x_j}-\frac{\partial \hat{A}_j}{\partial x_i}\Big) = i\hbar \epsilon_{ijk}\hat{H}_k
\end{equation}
In the above, we have $\hat{A}$ be a function of $\hat{x}$, which ensures commutation of the vector potential with itself (eq. \ref{eq:feynmanpositionassumption}). Our magnetic field can therefore be shown to be

\begin{equation}\label{eq:rightmag}
    \hat{H}_l= -\frac{i}{2\hbar}\epsilon_{jkl}[{\hat{P}}_j,{\hat{P}}_k]
\end{equation}
which avoids all the before-mentioned pitfalls.

Next, we demonstrate that $\hat{H}_l$ is a function of $\hat{x}$ and not $\hat{P}$:

\begin{equation}\label{eq:theoremref}
    [\hat{x}_i,\hat{H}_l] = 0
\end{equation}
By substituting in eq. \ref{eq:rightmag} into the expression $[\hat{x}_i,\hat{H}_l]$, and then utilizing eq. \ref{eq:cannonicalcommutatorfeynmanimproved} and the Jacobi identity, we derive the above result. $\hat{H}$ would therefore only be a function of $\hat{x}$ and $t$ \citep{Dyson}, and all components of $\hat{H}$ would commute with each others (this is also obvious when we plug eq. \ref{eq:rightmag} into $[\hat{H}_k,\hat{H}_l]$ and observe from symmetry that it must equal zero).

Next, using the Jacobi identity again, we prove no magnetic monopoles. The Jacobi identity for different $\hat{P}_l$ components:
\begin{equation}
    [\hat{P}_l,\epsilon_{jkl}[\hat{P}_j,\hat{P}_k]] + [\hat{P}_j,\epsilon_{jkl}[\hat{P}_k,\hat{P}_l]] + [\hat{P}_k,\epsilon_{jkl}[\hat{P}_l,\hat{P}_j]]=0
\end{equation}
which immediately implies:
\begin{equation}
[\hat{P}_k,\hat{H}_k]=0
\end{equation}
which is equivalent to $div ~\hat{H} =0$, using the fact that $\hat{H}$ is a function of $\hat
x$.

To prove the next Maxwell equation \citep{Dyson}, we take the time derivative of eq. \ref{eq:rightmag}:

\begin{equation}
    \frac{\partial \hat{H}_l}{\partial t} + \frac{\partial \hat{H}_l}{\partial x_m}\dot{\hat{x}}_m = -\frac{i}{\hbar}\epsilon_{jkl}[{\dot{\hat{P}}}_j,{\hat{P}}_k]
\end{equation}
Substituting in eqs. \ref{eq:newtonssecondlaw} and \ref{eq:referto1} in the same manner as \citet{Dyson}, we get:

\begin{equation}\label{eq:referenceto4}
    \frac{\partial \hat{H}_l}{\partial t} + \frac{\partial \hat{H}_l}{\partial x_m}\dot{\hat{x}}_m = -\frac{i}{\hbar}\epsilon_{jkl}[\hat{E}_j, \hat{P}_k]-\frac{i}{\hbar}[\dot{\hat{x}}_k\hat{H}_l,\hat{P}_k] + \frac{i}{\hbar}[\dot{\hat{x}}_l\hat{H}_j,\hat{P}_j] 
\end{equation}
Since eq. \ref{eq:cannonicalcommutatorfeynmanimproved} implies that $ [\dot{\hat{x}}_i,\hat{P}_j] = -[\hat{x}_i,\dot{\hat{P}}_j]$, we evaluate the last two terms in the above expression to be:
$$[\dot{\hat{x}}_k\hat{H}_l,\hat{P}_k] - [\dot{\hat x}_l\hat H_j,\hat P_j] = -[\hat x_k,\dot{\hat P}_k]\hat H_l - [\dot{\hat P}_j,{\hat x}_l]\hat H_j - \dot{\hat x}_k[\hat P_k,\hat H_l] + \dot{\hat x}_l[\hat P_j,\hat H_j]$$
The third term on the left cancels with the second term on the right-hand side of eq. \ref{eq:referenceto4}. The last term must be zero by no magnetic monopoles.
Using eqs. \ref{eq:newtonssecondlaw} and \ref{eq:referto1} again, we evaluate the remaining terms to be:
$$ -[\hat x_k,\dot{\hat P}_k]\hat H_l - [\dot{\hat P}_j,{\hat x}_l]\hat H_j  = \epsilon_{juv}[\hat x_l,\dot{\hat x}_u]\hat H_v \hat H_j + \epsilon_{juv}\dot{\hat x}_u[\hat x_l,\hat H_v]\hat H_j -\epsilon_{kyz}[\hat x_k, \dot{\hat x}_y]\hat H_z\hat H_l -\epsilon_{kyz}\dot{\hat x}_y[\hat x_k,\hat H_z]\hat H_l$$
Using eq. \ref{eq:theoremref} and symmetry, we can see that all the terms in the above right-hand side expression must be equivalent to zero. If we finally put everything together, we achieve our second Maxwell equation \citep{Dyson}:

\begin{equation}
    \frac{\partial \hat{H}_l}{\partial t} = \epsilon_{jkl}\frac{\partial \hat{E}_j}{\partial x_k}
\end{equation}
This concludes the relativistic Feynman proof with classical commutators. It is relativistic as we utilize the relativistic momentum operator $\hat{p}_j$ with minimal coupling. No nonrelativistic assumptions were introduced, unlike the original proof. No Galilean assumptions appear in the proof, contrary to the usual understanding of Feynman's proof (as presented by \citealt{Dyson}). Of interest, \citet{Sakurai} demonstrate eq. \ref{eq:referto1} by using the non-relativistic Hamiltonian, minimal coupling, and the Heisenberg equations of motion (see pp. 126 eq. 2.347).

One interesting aspect of Feynman's derivation is that 
charge and charge current seem to be afterthoughts, and the fields appear primary. It is interesting to consider the perspective of charge and current as being emergent from physically real fields of potential and momentum, instead of the fields being byproducts of charge. The usage of a wave commutator on relativistic Newton's Laws to derive wave equations of force appears less of an anomaly given the main text of this paper. Maxwell’s equations emerge naturally from a wave-based operator framework. 

\bibliographystyle{apalike}
\bibliography{hilbert_space}

\end{document}